\documentclass[11pt]{article}
\usepackage{epstopdf}                                                          
\usepackage[a4paper,hmarginratio=1:1,vmarginratio=2:3,
  totalwidth=15.2cm,totalheight=22.55cm]{geometry}
\usepackage{bm,epstopdf,epsfig,amsmath,amssymb,
amsfonts,colordvi,wrapfig,comment,cancel,verbatim,slashed}
\usepackage{graphicx,graphics}
\usepackage[font=md,captionskip=8pt]{subfig}
\usepackage[usenames,dvipsnames]{color}
\usepackage[noadjust]{cite}
\usepackage{xcolor} 
\usepackage[utf8]{inputenc}
\usepackage{setspace}

\usepackage[utf8]{inputenc}

\newcommand{\w}{\omega}  

\newcommand{\tGamma}{\tilde\Gamma}
\allowdisplaybreaks
\newcommand{\q}{\alpha}

\newcommand{\cO}{{\cal O}}

\newcommand{\cT}{{\cal T}}

\newcommand{\tr}{\text{tr}}

\newcommand{\be}{\begin{equation}}
\newcommand{\ee}{\end{equation}}
\newcommand{\bea}{\begin{eqnarray}}
\newcommand{\eea}{\end{eqnarray}}
                
\newcommand{\ra}{\rightarrow}  
\newcommand{\Ra}{\Rightarrow}

\newcommand{\baa}{\begin{array}}
\newcommand{\eaa}{\end{array}}

\long\def\symbolfootnote[#1]#2{\begingroup
\def\thefootnote{\fnsymbol{footnote}}\footnote[#1]{#2}\endgroup}

\begin{document} 
\begin{flushright}
\end{flushright}

\thispagestyle{empty}
\vspace{3cm}
\begin{center}


{\Large \bf  Weyl gauge invariant DBI action in conformal geometry}

 \vspace{2cm}
 
 {\bf D. M. Ghilencea}
 \symbolfootnote[1]{E-mail: dumitru.ghilencea@cern.ch}

\bigskip 

{\small Department of Theoretical Physics, National Institute of Physics
 \smallskip 

 and  Nuclear Engineering (IFIN), Bucharest, 077125 Romania}
\end{center}

\medskip

\begin{abstract}
\begin{spacing}{0.99}
  \noindent
We construct the analogue of the  Dirac-Born-Infeld (DBI) action in Weyl
conformal geometry in $d$ dimensions and obtain  a general theory of gravity
with Weyl gauge symmetry of dilatations (Weyl-DBI). This is done in the
{\it Weyl gauge covariant} formulation of conformal geometry in $d$ dimensions,
suitable for a gauge theory, in which  this geometry is {\it metric}.
The Weyl-DBI action is a special gauge theory
  in that it has the same gauge invariant expression with
{\it dimensionless} couplings in any dimension $d$, with no need for a  UV regulator
(be it a DR subtraction scale, field or higher derivative operator)
for which reason we argue it is Weyl-anomaly free. For $d=4$ dimensions, the leading order
of a series expansion of  the Weyl-DBI action recovers the  gauge
invariant Weyl quadratic gravity action associated to this geometry,
that is Weyl anomaly-free; this  is broken spontaneously and Einstein-Hilbert
gravity is recovered in the broken phase, with $\Lambda>0$. All the remaining
terms of this series expansion are of non-perturbative nature
but can, in principle,  be recovered  by (perturbative) quantum corrections
in  Weyl quadratic gravity in $d=4$ in a  gauge invariant (geometric)
regularisation,  provided by the Weyl-DBI action.
If the Weyl gauge boson is not dynamical the Weyl-DBI action recovers
in the leading order the conformal gravity action. All fields and scales 
have {\it geometric origin}, with no added matter, scalar field
compensators or UV regulators.
\end{spacing}
\end{abstract}

\newpage

\section{Motivation}

In this work we construct the analogue of the Dirac-Born-Infeld (DBI)
action \cite{BI,D,Sorokin,Gibbons,DeserGibbons}
in Weyl conformal geometry \cite{Weyl1,Weyl2,Weyl3} in $d$ dimensions.
The motivation for  Weyl geometry is  that it is the underlying geometry of
an ultraviolet completion of Einstein-Hilbert
gravity and SM in a {\it gauge theory} of the Weyl group (of dilatations and
Poincar\'e symmetry), as explained below. The action we construct
is relevant for gravity theories based on conformal geometry.

Let us motivate our interest in Weyl conformal geometry.
First, the quadratic gravity action defined by  this geometry 
is a gauge theory of the Weyl group;  in the {\it absence} of matter
the gauged dilatations  symmetry is broken by a Stueckelberg mechanism
in which the Weyl gauge field ($\w_\mu$) becomes massive and
decouples at low scales and Riemannian geometry and Einstein-Hilbert gravity
are nicely recovered \cite{Ghilen0,SMW}.
The Planck scale is generated by the dilaton propagated by
the $\hat R^2$ term. Further, Weyl conformal geometry admits
a {\it Weyl gauge covariant} formulation  \cite{DG1,Dirac}
that is automatically  {\it metric}. As a result of this and
contrary to a long-held view \cite{Weyl1}, there is no second clock effect:
under parallel transport the length of a vector and clock rates are invariant {\it if
  the transport respects Weyl gauge covariance}, as it should 
in order to be physical  \cite{non-metric} (Appendix B), \cite{CDA,Lasenby}.
Briefly, an  Einstein-Hilbert gravity is recovered in the broken phase of the
(gauge theory of) Weyl quadratic gravity. All fields and scales have geometric origin,
with no added matter or Weyl scalar field compensators, etc.

Adding matter is immediate: the SM with a vanishing Higgs mass parameter
admits a natural and truly minimal embedding in conformal geometry
\cite{SMW} {\it without} new degrees of freedom beyond those of SM and Weyl geometry!
In the limit of a vanishing Weyl gauge current, $\w_\mu$ becomes ``pure gauge'' and
the Weyl quadratic gravity  action reduces to a conformal gravity action
(i.e. Weyl-tensor-squared) \cite{non-metric,GH}, which is thus less general.
Successful Starobinsky-Higgs inflation is possible \cite{WI3,WI1,WI2}
being a gauged version of
Starobinsky inflation \cite{Starobinsky}; good fits for the galaxies
rotation curves suggest a geometric solution for dark matter
associated to $\w_\mu$ \cite{Harko};  black hole solutions
were studied in \cite{Harko2}. The presence of $\w_\mu$  seems
necessary for geodesic completeness of Weyl geometry \cite{Ohanian,Ehlers}.
Weyl geometry seems also relevant for the boundary CFT of  the AdS/CFT holography
 \cite{Jia,Ciambelli}.

 Using the Weyl gauge covariant  formulation of Weyl conformal geometry, which renders
 it {\it metric},  one shows  that the gauged dilatations symmetry of Weyl quadratic
gravity is actually maintained at the quantum level in $d$ dimensions and hence this symmetry
is Weyl anomaly-free \cite{DG1} - as it should be for  a consistent (quantum) gauge symmetry.
This differs from gravity actions in Riemannian geometry
where the well-known Weyl anomaly is present \cite{Duff,Duff2,Duff3,Deser1976}.
The absence here of  Weyl anomaly 
is due to Weyl gauge covariance in $d$ dimensions of both the 
Weyl term $\hat C_{\mu\nu\rho\sigma}^2$ {\it and} the Chern-Euler-Gauss-Bonnet term $\hat G$
in the action\footnote{This covariance enables a Weyl gauge invariant (geometric)
regularisation of the action \cite{DG1}.} {\it as well as} to an
additional  dynamical degree of freedom (``dilaton'' or, more exactly,
would-be Goldstone of gauged dilatations, $\phi$), compared to a Riemannian case.
Weyl anomaly is recovered in the  broken phase \cite{DG1} after $\phi$
eaten by $\w_\mu$  decouples  with massive $\w_\mu$, and
Weyl geometry (connection) becomes Riemannian.
Briefly, Weyl conformal geometry is more fundamental:
conformal geometry is a (metric) gauge covariant
extension of Riemannian geometry with respect to the extra
gauged dilatation symmetry of the Weyl group \cite{DG1}.
This way conformal geometry becomes the underlying geometry of a
unified anomaly-free  gauge theory of gravity and SM \cite{review}.

Here we construct an analogue of  DBI action of Weyl gauge symmetry
in  Weyl conformal geometry in $d$ dimensions (Weyl-DBI).
This action is  special  because it has the same Weyl
gauge invariant expression with dimensionless couplings
in any dimension $d$ i.e. it
has {\it no UV regulator} (be it an extra field, higher derivative operator
or subtraction  scale) usually required by the analytic
continuation of familiar gauge theories (e.g. Yang-Mills);
for this reason  it is, arguably,  Weyl anomaly-free if coupled to matter
in  Weyl gauge invariant way.

The Weyl-DBI action generalizes the above $d=4$ Weyl quadratic gravity
of conformal geometry,  and automatically provides it with a
 Weyl gauge invariant {\it geometric}
regularisation in $d\!=\!4-2\epsilon$ with the  scalar curvature ($\hat R$)
playing the actual role of  regulator. The leading order of a series
expansion of the Weyl-DBI action recovers  exactly
the Weyl quadratic gravity gauge theory, while all subleading terms
have a  non-perturbative  structure (suppressed by powers of $\hat R$);
nevertheless, some of these can be generated perturbatively, at quantum level,
by Weyl quadratic gravity in $d=4$ in the Weyl gauge invariant, geometric
regularisation. The above interesting connection
of  Weyl-DBI gauge theory to the realistic (gauge theory of) Weyl quadratic gravity,
to which it provides a generalization and embedding,  motivated the present
study.

\section{Weyl conformal geometry and its gravity}\label{1}

We first review Weyl conformal geometry and its associated quadratic gravity action;
this action is  a gauge theory of the Weyl group of dilatations and Poincar\'e symmetry.
We review  this geometry in the Weyl gauge covariant and metric
formulation of \cite{DG1,Dirac} and the  breaking of this symmetry  \cite{SMW,Ghilen0}.
For the original work  on Weyl geometry see
\cite{Weyl1,Weyl2,Weyl3}; for a historical review and references,
but in a non-covariant, non-metric formulation, see \cite{Scholz}.

Weyl  geometry is {\it defined} by classes of equivalence
$(g_{\alpha\beta}, \w_\mu$) of the metric ($g_{\alpha\beta}$)
and Weyl gauge field of dilatations ($\w_\mu$),  related by a Weyl gauge
transformation shown here in $d$ dimensions\footnote{Our conventions are
as in the book  \cite{b1} with $(+,-,-,-)$ for the metric.}
\smallskip
\bea\label{WGS}
 &\quad&
 g_{\mu\nu}^\prime=\Sigma^q 
 \,g_{\mu\nu},\qquad
 \w_\mu'=\w_\mu -\frac{1}{\q}\, \partial_\mu\ln\Sigma, 
\qquad
\sqrt{g'}=\Sigma^{q d/2} \sqrt{g}.
\eea

\medskip\noindent
Here $q$ is the Weyl charge of the metric; various conventions exist
for the charge normalization: $q\!=\!2$, etc; here we keep $q$ arbitrary;
$\alpha<1$ is the  gauge coupling of dilatations.
If scalars ($\phi$) or fermions ($\psi$) exist, then (\ref{WGS}) is completed by
\footnote{{If $q=2$, $d=4$, Weyl charges are the usual
inverse mass dimensions of the fields: $q_\phi=-1$, $q_\psi=-3/2$. }}
\smallskip\bea\label{wgs}
 &\quad & \phi' = \Sigma^{q_\phi} \phi, 
\quad
\qquad \psi'=\Sigma^{q_\psi}\,\psi,
 \qquad q_\phi=-\frac{q}{4} (d-2),
\qquad  q_\psi=-\frac{q}{4} (d-1).
\eea

\medskip
{Transformations (\ref{WGS}), (\ref{wgs}) define
  the  Abelian gauged dilatation $D(1)$ or {\it Weyl gauge symmetry}
  of this geometry; this symmetry
  extends the usual   (local) Weyl symmetry, by the presence
of a Weyl gauge boson $\omega_\mu$;
 ($\w_\mu=0$ or  `pure gauge' in a local Weyl symmetry case).
The field $\omega_\mu$ together with the metric $g_{\mu\nu}$ and
symmetry (\ref{WGS}) are part of Weyl geometry definition that is completed
by a non-metricity condition ($\tilde\nabla_\mu g_{\alpha\beta}\!\not=\! 0$)
which is}
\smallskip 
\bea\label{tildenabla}
\tilde\nabla_\lambda g_{\mu\nu}=- q \,\alpha \omega_\lambda g_{\mu\nu},
\qquad \textrm{where}\qquad
\tilde\nabla_\lambda g_{\mu\nu}=\partial_\lambda g_{\mu\nu}-\tilde\Gamma^\rho_{\lambda\mu} g_{\rho\nu}
-\tilde \Gamma^\rho_{\lambda\nu} g_{\rho\mu}.
\eea
Here, the Weyl connection $\tilde\Gamma_{\mu\nu}^\lambda$ is assumed symmetric
($\tilde\Gamma_{\mu\nu}^\lambda=\tilde\Gamma_{\nu\mu}^\lambda$). $\tilde\Gamma_{\mu\nu}^\lambda$ is
found by direct calculation or 
by a covariant derivative substitution of $\partial_\mu$: 
$\tilde \Gamma_{\mu\nu}^\lambda=\Gamma_{\mu\nu}^\lambda\big
\vert_{\partial_\mu \ra \partial_\mu+q\,\alpha \omega_\lambda}$; $\Gamma_{\mu\nu}^\lambda$ 
is the Levi-Civita (LC) connection with
$\Gamma_{\mu\nu}^\rho=(1/2) g^{\rho\lambda} (\partial_\mu g_{\nu\lambda}
 +\partial_\nu g_{\mu\lambda}- \partial_\lambda g_{\mu\nu})$. One finds
\medskip
\bea\label{tG}
\tilde \Gamma_{\mu\nu}^\lambda
=\Gamma_{\mu\nu}^\lambda
+\tilde\alpha\big[\delta_\mu^\lambda \omega_\nu +
  \delta_\nu^\lambda \omega_\mu-g_{\mu\nu} \omega^\lambda\big],\qquad
{\rm with\,\,\,notation}\qquad
\tilde\alpha\equiv \alpha\, q/2.
\eea

\medskip\noindent
{Therefore
  $\omega_\mu$ is part of the  Weyl geometry connection $\tilde\Gamma$,
  hence it has  geometric origin.
 Since $\omega_\mu\propto \tilde\Gamma_{\mu\nu}^\nu-\Gamma_{\mu\nu}^\nu$,
   $\omega_\mu$ measures the (trace of) deviation of Weyl connection
  from Levi-Civita connection; if $\omega_\mu$ vanishes one recovers Riemannian
  geometry.}\footnote{
  Here  is another, physical motivation to consider Weyl conformal geometry.
  A dynamical $\omega_\mu$ is needed in theories with
  local Weyl symmetry  in Riemannian geometry,
  to ensure the Einstein term and the dilaton kinetic term
  have correct signs (no ghost), with Planck scale generated by the dilaton vev.
  The action so obtained is a simple version (linear in $\hat R$)
of that in Weyl geometry shown later, see \cite{Lee} (section~2).
  This way one extends local Weyl symmetry to Weyl gauge symmetry (\ref{WGS})
  and brings us effectively to Weyl geometry.}

The  Weyl connection ($\tilde \Gamma$) is invariant under (\ref{WGS}).
One usually defines a Riemann tensor in Weyl geometry by the standard formula
of Riemannian case but with $\Gamma$ replaced by $\tilde\Gamma$:
\medskip
\bea\label{Rie}
\tilde R^\lambda_{\,\,\,\,\mu\nu\sigma}
&=&
\partial_\nu\tGamma^\lambda_{\sigma\mu}
-\partial_\sigma\tGamma_{\nu\mu}^\lambda
+\tGamma_{\nu\rho}^\lambda\,\tGamma^\rho_{\sigma\mu}
-\tGamma_{\sigma\rho}^\lambda\,\tGamma^\rho_{\nu\mu}.
\eea

\medskip\noindent
This can  be expressed in terms of the Riemann tensor of
Riemannian geometry, using (\ref{tG}).
 $\tilde\Gamma$ is invariant under (\ref{WGS}), then
$\tilde R^\lambda_{\,\,\mu\nu\sigma}$ and  the Ricci tensor of Weyl geometry
$\tilde R_{\mu\nu}=\tilde R^\lambda_{\,\,\,\mu\lambda\nu}$ are also invariant.
The only issue is that Weyl geometry
being non-metric i.e. $\tilde\nabla_\mu g_{\alpha\beta}\not=0$,
to do  calculations one must go to the (metric)
Riemannian picture. This complicates significantly the calculations
since Riemannian geometry (connection) does not have symmetry (\ref{WGS}).

This (apparent) non-metricity is however an artefact of this formulation
which does not maintain manifest Weyl gauge covariance of e.g.
the derivatives of  curvature tensors and scalar, required in a gauge theory of (\ref{WGS}).
As shown in \cite{DG1} Weyl geometry admits however another (equivalent)
formulation which is
{\it Weyl gauge covariant} in which this geometry is automatically {\it metric}\,
(see  \cite{CDA,CA} for an in-depth analysis of the equivalent
formulations). This is  important since it allows  a) the usual 
gauge theory covariant approach for its associated quadratic gravity action and
b):  it enables us to do calculations directly in Weyl geometry (e.g.
anomaly calculation \cite{DG1}) like in (metric) Riemannian geometry,
hence, no need to go to a Riemannian picture. We  summarize
below this  formulation \cite{DG1,CDA}.

To find a  Weyl gauge covariant and metric formulation, recall 
 that $(\tilde\nabla_\lambda +q \, \alpha\w_\lambda) g_{\mu\nu}=0$ with $q$  the Weyl
charge of $g_{\mu\nu}$: this suggests that for any given tensor $T$ of charge $q_T$, 
in particular $g_{\mu\nu}$,  with $T^\prime=\Sigma^{q_T} T$ one defines
a new differential operator $\hat \nabla_\mu$ to replace
$\tilde\nabla_\mu$
\medskip
\bea\label{qq}
\hat \nabla_\mu T
=\tilde\nabla_\mu\Big\vert_{\partial_\mu\ra \partial_\mu + q_T\alpha \w_\mu} T
\equiv (\tilde\nabla_\mu +q_T\,\alpha\, \w_\mu) T\qquad
\Ra\qquad \hat\nabla_\mu' T'=\Sigma^{q_T}\, \hat\nabla_\mu T.
\eea

\medskip\noindent
Hence,  $\hat\nabla_\mu$ transforms covariantly under (\ref{WGS}),
as seen by using
that $\tilde\Gamma$ is invariant. Eq.(\ref{qq}) simply introduces a Weyl gauge covariant
$\hat\nabla_\mu$  by  covariantising the partial derivative $\partial_\mu$
in $\tilde\nabla_\mu$: $\partial_\mu\ra\partial_\mu\! +\!\rm{charge} \times\alpha\,\w_\mu$.
The theory is now  {\it metric} with respect to $\hat \nabla_\mu$, since
$\hat\nabla_\mu g_{\alpha\beta}=0.$

One  then defines  new Riemann and Ricci tensors of Weyl geometry (with a `hat') 
using the new differential operator $\hat\nabla_\mu$ in the commutator that defines
the Riemann tensor \cite{CDA}
\medskip
\bea\label{rrr}
[\hat\nabla_\nu,\hat\nabla_\sigma] v^\lambda=\hat R^{\lambda}_{\,\,\mu\nu\sigma}\, v^\mu
\eea

\medskip\noindent
$v^\rho$ is a vector of vanishing Weyl charge on tangent space.
With $\hat R_{\alpha\mu\nu\sigma}\!=g_{\alpha\lambda} \hat R^\lambda_{\,\,\,\mu\nu\sigma}$ then
\medskip
\bea\label{ss1}
\hat R_{\alpha\mu\nu\sigma}
&=&
R_{\alpha\mu\nu\sigma}
+\tilde\alpha\Big\{g_{\alpha\sigma} \nabla_\nu\w_\mu- g_{\alpha\nu} \nabla_\sigma \w_\mu
-g_{\mu\sigma}\nabla_\nu \w_\alpha +g_{\mu\nu} \nabla_\sigma\w_\alpha
\Big\}
\nonumber\\
&+&\tilde\alpha^{2} \Big\{
\w^2 (g_{\alpha\sigma} g_{\mu\nu}-g_{\alpha\nu} g_{\mu\sigma} )
+\w_\alpha \,(\w_\nu g_{\sigma\mu}-\w_\sigma g_{\mu\nu})
+\w_\mu (\w_\sigma g_{\alpha\nu} -\w_\nu g_{\alpha\sigma})\Big\}\quad
\eea

\medskip\noindent
where $R_{\alpha\mu\nu\sigma}$ is that of  Riemannian geometry and so is $\nabla_\mu$
acting with
LC connection: $\nabla_\mu\omega_\nu=\partial_\mu \omega_\nu -\Gamma_{\mu\nu}^\rho\omega_\rho$.
The relation to (\ref{Rie}) is
$\hat R^\lambda_{\,\,\mu\nu\sigma}=\tilde R^\lambda_{\,\,\mu\nu\sigma}- \tilde\alpha
\delta_\mu^\lambda \hat F_{\nu\sigma}$.
Also we have $\hat F_{\mu\nu}=\hat\nabla_\mu\w_\nu-\hat\nabla_\nu\w_\mu
=\partial_\mu\w_\nu-\partial_\nu\w_\mu=\nabla_\mu\w_\nu-\nabla_\nu\w_\mu
= F_{\mu\nu}$, since $\tilde\Gamma$ and $\Gamma$ are symmetric in their lower indices.
Like $\tilde R^\lambda_{\,\,\mu\nu\sigma}$,  $\hat R^\lambda_{\,\,\mu\nu\sigma}$
is  Weyl gauge invariant, too.
The Ricci tensor in Weyl geometry
is then $\hat R_{\mu\sigma}=\hat R^\lambda_{\,\,\,\,\mu\lambda\sigma}$
giving
\be\label{ss2}
\hat R_{\mu\sigma}=
R_{\mu\sigma} 
\!+\!\tilde\alpha\, 
\Big[\frac12 (d-2) F_{\mu\sigma}-(d-2)\nabla_{(\mu} \omega_{\sigma)}
 - g_{\mu\sigma} \nabla_\lambda\omega^\lambda\Big]
\!+\!\tilde\alpha^{2}  (d-2) (\omega_\mu\omega_\sigma -g_{\mu\sigma} \omega_\lambda\omega^\lambda)
\ee

\medskip\noindent
with $R_{\mu\nu}$ the Ricci tensor in Riemannian geometry.
Note that $\hat R_{\mu\nu}-\hat R_{\nu\mu}=\tilde\alpha (d-2) F_{\mu\nu}$.

Further,  the  Weyl scalar curvature  $\hat R$ of Weyl geometry is
\medskip
\bea\label{Rs}
\hat R=g^{\mu\sigma}\hat R_{\mu\sigma}=R-2 (d-1)\, \tilde\alpha \, \nabla_\mu \omega^\mu 
-(d-1) (d-2) \,\tilde\alpha^{2} \omega_\mu \omega^\mu,
\eea

\medskip\noindent
in terms of scalar curvature $R$ of Riemannian geometry, $R=g^{\mu\nu} R_{\mu\nu}$.
The  Weyl tensor in Weyl geometry  associated to $\hat R_{\mu\nu\rho\sigma}$
 is then (with $\hat C^\mu_{\,\,\,\nu\mu\sigma}=0$) \cite{DG1}
 \medskip
 \bea\label{tc}
\hat C_{\alpha\mu\nu\sigma}=C_{\alpha\mu\nu\sigma},
\eea

\medskip\noindent
with $C_{\alpha\mu\nu\sigma}$ the Riemannian geometry counterpart.
So in this formulation the Weyl tensor has the same expression in both geometries.
Finally, there is the Chern-Euler-Gauss-Bonnet term $\hat G$ (hereafter `Euler term') which
in the metric (`hat') formulation is \cite{DG1}
\medskip
\bea\label{hatG}
\hat G= \hat R_{\mu\nu\rho\sigma} \hat R^{\rho\sigma\mu\nu}
- 4 \hat R_{\mu\nu} \hat R^{\nu\mu} +\hat R^2,
\eea

\medskip\noindent
and is a total derivative in $d=4$.
Note the position of the summation indices.

With these formulae one easily finds that we have the following invariants
under (\ref{WGS})
\medskip
\bea\label{inv}
\hat R_{\mu\nu}^\prime=\hat R_{\mu\nu},\quad
\hat R^\prime \, g_{\mu\nu}^\prime=\hat R\, g_{\mu\nu}, \quad
\hat F_{\mu\nu}^\prime=\hat F_{\mu\nu},
\eea

\medskip\noindent
and  manifest {\it Weyl gauge covariance } of the fields and of their derivatives
\medskip
\bea
\label{WGS3}
&& \hat R^\prime=\Sigma^{-q} \hat R,
\nonumber\\[5pt]
&& X^\prime=\Sigma^{-2 q} X, \qquad
X  =\hat R_{\mu\nu\rho\sigma}^2, \, \,\hat R_{\mu\nu}^2,\,\,\hat R^2,
 \,\, \hat C_{\mu\nu\rho\sigma}^2,\,\, \hat G,
\,\, \hat F_{\mu\nu}^2,
\nonumber\\[5pt]
&&\hat\nabla_\mu' \hat R^\prime=\Sigma^{-q}\hat \nabla_\mu \hat R,\qquad
\hat\nabla^\prime_\mu\hat\nabla^{\prime}_\nu \hat R'
=\Sigma^{-q} \hat \nabla_\mu\hat\nabla_\nu \hat R, \quad 
\hat\nabla_\alpha' \hat R_{\mu\nu}'=\hat\nabla_\alpha \hat R_{\mu\nu},\,\,\,
\text{etc.}
\eea

\medskip\noindent
It is important to note
that the Euler term $\hat G$ is  Weyl gauge covariant in $d$ dimensions,
a property specific to Weyl conformal geometry that is not true in Riemannian case!
This property is important since it ensures Weyl quadratic gravity is
Weyl anomaly-free \cite{DG1}.

With formulae (\ref{WGS3}), Weyl geometry can  be regarded as a
{\it covariantised version} of Riemannian geometry with
respect to Weyl gauge symmetry (\ref{WGS})
that is also metric ($\hat \nabla_\mu g_{\alpha\beta}\!=\!0$).

Let us  present two identities in Weyl conformal
geometry used later, that generalize  those of Riemannian geometry.
In the  metric Weyl gauge covariant formulation one shows after a long
algebra \cite{DG1} (Appendix)
\medskip
\bea\label{g1}
\hat C_{\mu\nu\rho\sigma}^2= \hat R_{\mu\nu\rho\sigma} \hat R^{\rho\sigma\mu\nu}
-\frac{4}{d-2} \hat R_{\mu\nu} \hat R^{\nu\mu} +\frac{2}{(d-1)(d-2)}\hat R^2.
\eea

\medskip\noindent
With (\ref{hatG}), we  express the Ricci tensor-squared in terms of the Weyl tensor:
\medskip
\bea\label{g2}
\hat R_{\mu\nu} \hat R^{\nu\mu}=\frac{d-2}{4 (d-3)} \big( \hat C_{\mu\nu\rho\sigma}^2 -\hat G\big)
+\frac{d}{4 (d-1)} \hat R^2.
\eea

\medskip\noindent
The last two relations are ``covariantised'' versions of the similar ones in Riemannian
geometry with respect to the Weyl gauge symmetry \cite{DG1,review}.
This ends our review  on conformal geometry
in a metric, Weyl gauge covariant formulation.

The most general gravity action in Weyl conformal geometry
is quadratic in curvature. In $d=4$  this action  is shown below in a
basis of independent operators \cite{Weyl2}
\medskip
\bea\label{WA}
S_{\bf w}=\int d^4x \, \sqrt{g} \,\,\Big[\, \frac{1}{4!\,
    \xi^2} \,\hat R^2 -\frac14 \,\hat F_{\mu\nu}^2
  -\frac{1}{\eta^2}\, \hat C_{\mu\nu\rho\sigma}^2  +\frac{1}{\rho}\,\hat G\, \Big]
\eea

\medskip\noindent
where $\xi, \eta$ are perturbative couplings ($<1$) and $g=\vert \det g_{\mu\nu}\vert$;
{$\rho$ is here an arbitrary coupling that can be chosen at will,
since the Euler term $\hat G$ does  not affect the equations of motion in  $d=4$ dimensions,
but it does so in  $d=4-2\epsilon$ when $\rho$ becomes relevant.}

Given (\ref{WGS3}), action (\ref{WA}) is invariant under (\ref{WGS}) and is
generated by Weyl conformal geometry alone which can thus  be seen as a gauge theory of
of dilatations. Higher dimensional  operators suppressed by some
mass scale are not allowed in (\ref{WA})  since, if present,
such scale would break symmetry (\ref{WGS}).

The  Weyl gauge symmetry of action (\ref{WA}) is broken spontaneously 
by a Stueckelberg mechanism as first shown  in \cite{Ghilen0} (with applications in
\cite{SMW,DG1,CDA,non-metric,WI3,WI1,WI2,Harko,Harko2}). Let us detail this.
First,  one linearises the quadratic term in (\ref{WA}) with the aid
of a scalar field $\phi$ by replacing $\hat R^2\ra -2 \phi^2\hat R-\phi^4$ in $S_{\bf w}$; the
solution of the equation of motion of $\phi$ is then $\phi^2=-\hat R$ ($\hat R<0$)
which, when replaced back
in the new action, recovers (\ref{WA}), hence the actions before and after replacement
are classically equivalent. One then  writes the new $S_{\bf w}$
in  a Riemannian notation using  eqs.(\ref{Rs}), (\ref{tc}).
Next, there is a  Stueckelberg mechanism \cite{Stueckelberg} in new $S_{\bf w}$,
where $\w_\mu$ is eating the (derivative of the)
 dilaton\footnote{Notice that the field $\ln\phi$ transforms with
 a shift under (\ref{WGS}).} $\ln\phi$, to become massive.
When $\phi$ acquires a vev, one obtains \cite{SMW,Ghilen0,DG1}
from (\ref{WA}) the Einstein-Proca action for
 massive $\w_\mu$, a positive cosmological constant and a Weyl-tensor-squared term.
 The Weyl gauge symmetry is broken, massive $\w_\mu$ now
 decouples and below its mass Weyl geometry
 (connection) becomes Riemannian  geometry (connection), respectively,
so  $\tilde\Gamma\ra \Gamma$, see (\ref{tG}).
 In a {\it Riemannian} notation the broken phase of  $S_{\bf w}$ is
 \cite{SMW,Ghilen0,DG1} (see e.g. eq.18 in \cite{SMW})
\medskip
\be
\label{EP}
S_{\bf w}=\int d^4x
\sqrt{g}  \,\Big[- \frac12\, M_p^2 \, R
 +\frac12 m_\omega^2\, \w_\mu \, \w^{\mu}
 - \Lambda\, M_p^2
 -\frac{1}{4} \, F_{\mu\nu}^{ 2}-\frac{1}{\eta^2}\,  C_{\mu\nu\rho\sigma}^2
 \Big],
\ee

\medskip\noindent
where we denoted
\medskip
\bea\label{la}
\Lambda\equiv \frac14\,\langle\phi\rangle^2,\qquad
M_p^2\equiv \frac{\langle\phi^2\rangle}{6\,\xi^2},\qquad
m_\omega^2\equiv \frac32 \alpha^2\,q^2\,M_p^2,
\eea

\medskip\noindent
with $M_p$ and $\Lambda$  identified with Planck
scale and cosmological constant, respectively. $\Lambda$ is small 
because the (dimensionless) gravitational coupling is weak:  $\xi\ll 1$.
For a FLRW metric, one can show \cite{GH} that on the ground state $\Lambda=3 H_0^2$
and $\hat R=-12 H_0^2$  consistent with our convention
$\hat R\!<\!0$ ($H_0:$ Hubble constant).
Apart from  the $C_{\mu\nu\rho\sigma}^2$  term\footnote{The mass of
spin-two state due to  $C_{\mu\nu\rho\sigma}^2$
is $m\sim \eta M_P$ so for $\eta\sim 1$ this state decouples below $M_P$.} ($\eta\leq 1$),
Einstein-Hilbert action is recovered in the broken phase (\ref{EP}) of
gauge theory  (\ref{WA}) and  massive $\w_\mu$  can now decouple; $m_\w$ 
is between $1$ TeV ($\alpha\!\ll \!1$) and $M_p$ ($\alpha\!\sim\! 1$) \cite{Latorre,SMW}.
{For later use, a phenomenologically viable choice of couplings is then  e.g.
$\xi\ll \eta <1$ and $\alpha\sim 1$.}

What happens at quantum level?
To ensure that this (quantum) gauge symmetry is not anomalous  one
requires  first a regularisation  that preserves the Weyl gauge symmetry.
This is  possible  \cite{DG1} due to the
Weyl gauge covariance of both $\hat R$ and $\hat G$ in particular,  discussed above.
An analytic  continuation to $d=4-2\epsilon$ dimensions is then
\medskip
\bea\label{WAd}
S_{\bf w}=\int d^dx \, \sqrt{g} \,\,(\hat R^2)^{(d-4)/4}\,\Big[\, \frac{1}{4!\,
    \xi^2} \,\hat R^2 -\frac14 \,\hat F_{\mu\nu}^2
  -\frac{1}{\eta^2}\, \hat C_{\mu\nu\rho\sigma}^2  +\frac{1}{\rho}\,\hat G\, \Big]
\eea

\medskip\noindent
and $S_{\bf w}$  is invariant under (\ref{WGS}) with (\ref{WGS3}).
Since the Weyl gauge symmetry is manifest in $d$ dimensions this indicates
that $S_{\bf w}$ is anomaly free \cite{DG1}.  But the absence of Weyl anomaly
is here  more than a regularisation matter: it is due to the Weyl
gauge covariance of $\hat G$ that enables (\ref{WAd}) be invariant
but also  to the presence of an additional dynamical
``dilaton'' $\ln\phi$ (propagated by the  $\hat R^2$ term) that mixes with
the graviton \cite{Englert}\footnote{Conversely, in Riemannian case Weyl
anomaly signals the missing of such dynamical degree of freedom.}.
When $\ln\phi$  is eaten by $\w_\mu$ which becomes massive and decouples
(together with $\phi$), then Weyl geometry (connection) becomes Riemannian, see (\ref{tG}),
and Weyl anomaly is recovered in the broken phase \cite{DG1}.
{This ends our review of conformal geometry and its associated Weyl quadratic gravity in
  the Weyl gauge covariant, metric formulation.}

\section{Gauge invariant DBI action of Weyl conformal geometry}

\noindent {\bf $\bullet$ Weyl - DBI action  in $d$ dimensions}

\medskip\noindent
A natural question is whether  in Weyl conformal geometry 
there can exist an action more general than
(\ref{WA}) and (\ref{WAd})  that is Weyl gauge invariant.
This could be a more general candidate for
a Weyl anomaly-free (quantum) gauge theory of scale invariance that may be
physically relevant.
The answer  is given by the analogue of the DBI action\footnote{For some other
models of DBI action applied to gravity see for example
\cite{Deser,Vollick1,Vollick2,R1,R2,R3,R4}.}
for the Weyl gauge  symmetry $D(1)$  in Weyl conformal geometry.
In this section we discuss such DBI action in $d$ dimensions and consider some limits,
including $d=4$.

From (\ref{inv}), each term
$\hat R\, g_{\mu\nu}$, $\hat R_{\mu\nu}$, $\hat F_{\mu\nu}$ is invariant under 
gauge symmetry (\ref{WGS}) in $d$ dimensions.
We can then construct a DBI-like action  in conformal geometry
in $d$ dimensions
\bea\label{dbi}
S_d=\int d^d\sigma\,
\Big\{-
\det\,[a_0 \,\hat R\, g_{\mu\nu}+a_1 \,\hat R_{\mu\nu}+ a_2 \,\hat F_{\mu\nu}]\,
\Big\}^{\frac{1}{2}}
\eea
where $a_0, a_1, a_2$ are   dimensionless constants. 
This gauge theory action has a very special feature: it is Weyl gauge invariant
 with {\it dimensionless} couplings in arbitrary $d$ dimensions!
 thus, it has  no need for  a UV regulator.
 We return to this issue shortly.

The action contains  higher-derivative terms
and differs from   usual DBI action \cite{DeserGibbons} where the metric 
is not multiplied by scalar curvature, demanded here by the symmetry.
Further\footnote{
We use $\hat R\!\not=\!0$ since in a leading order $\cO(X^3)$  found later
we recover (\ref{WA}) giving $\hat R\!=\!-\phi^2\!\not=\!0$, see (\ref{la}).}
\bea
S_d &=&
\int d^d\sigma
\sqrt{g}\,\, (a_0\,\hat R)^{d/2}\,\,
\Big\{  \det \,\big[\delta_\nu^\lambda+X^\lambda_{\,\,\,\nu}\,\big]\Big\}^{\frac12}
\eea
where
\bea\label{X}
X^\lambda_{\,\,\,\,\nu}= \frac{1}{a_0\, \hat R}\, g^{\lambda\rho} 
\big[a_1\,\hat R_{\rho\nu}+  a_2 \,\hat F_{\rho\nu}\,\big].
\eea
 Further, 
 \bea
 \sqrt{\det \big(1+X\big)}\!\!
 &=&\!\!
1+\frac12\, \tr X+\frac14\,
\Big[\frac12 (\tr X)^2-\tr X^2\Big]
\nonumber\\[4pt]
&+&\!\!\!\!
+\Big[\frac{1}{48} (\tr X)^3-\frac{1}{8} \tr X \,\tr X^2
  +\frac16\,\tr X^3\Big]
+\cO(X^4)
\eea
%
where the higher order terms include all combined powers
of  $\tr$  and $X$ and a sufficient condition for
a rapid  convergence of the  expansion is
$\vert a_i/a_0\vert\ll 1$. $i=1,2$.
Using the properties of $\hat R_{\mu\nu}$ and $\hat F_{\mu\nu}$
we find
\medskip
\bea\label{traces}
\tr X&=& \frac{a_1}{a_0},\qquad
\tr X^2 = \frac{1}{a_0^2\,\hat R^2} \Big[
 a_1^2\, \hat R_{\mu\nu}\hat R^{\nu\mu}
+a_2 \Big( a_2 + a_1\,\alpha\,q\,\frac{d-2}{2}\Big) \hat F_{\mu\nu} \hat F^{\nu\mu}\Big]
\nonumber\\[3pt]
\tr X^3 &= & \frac{1}{\hat R^3} \, \Big(\frac{a_1}{a_0}\Big)^3 \hat R^{\mu\sigma} \hat R_{\sigma\rho}\,
\hat R_{\nu\mu} \, g^{\nu\rho} + \cdots
\eea

\medskip\noindent
Next, use (\ref{g2}), to replace $\hat R_{\mu\nu} \hat R^{\nu\mu}$
in terms of the Weyl tensor. Bringing everything together, we find
$S_d$ in a basis of independent operators:
\medskip
\bea\label{S3}
S_d=\int d^d\sigma \sqrt{g}\,\,  (\hat R^2)^{d/4-1}\, \Big[\,
  \frac{1}{4! \,\xi^2}  \,\hat R^2
    -\frac{1}{\zeta}
    \hat F_{\mu\nu} \, \hat F^{\mu\nu}
-\frac{1}{\eta^2}
  \hat C_{\mu\nu\rho\sigma}\hat C^{\mu\nu\rho\sigma}
    +\frac{1}{\eta^2}\, \hat G +\cO(X^3)\Big]
\eea

\medskip\noindent
where we denoted
\medskip
\bea\label{couplings}
\frac{1}{4!\, \xi^2}&=&
\Big[\, a_0^2+ \frac12 \,a_0\,a_1+ \,a_1^2\,\frac{d-2}{16 \,(d-1)}\,\Big]\,a_0^{\frac{d}{2}-2}\,;
\nonumber\\[4pt]
\frac{1}{\eta^2}&= &
\frac{1}{16} \frac{d-2}{(d-3)}\,a_1^2 \,a_0^{\frac{d}{2}-2},\quad
\frac{1}{\zeta}
=-\frac14 \, a_2  \Big[a_2 + a_1\,\alpha\,q \frac{d-2}{2}\Big]\,a_0^{\frac{d}{2}=2}.
\eea

\medskip
{The leading order of the Weyl-DBI action,  eq.(\ref{S3}),
  recovered all the terms of the Weyl quadratic gravity action 
with Weyl gauge symmetry in $d$ dimensions, shown in  eq.(\ref{WAd})!
Even the couplings of these two actions can be equal, 
if   $\rho$, which is actually arbitrary in eq.(\ref{WA}), is set to $\rho=\eta^2$;
if so,  a solution  $a_{0,1,2}$ to the system of eqs.(\ref{couplings})
(with $\zeta=4$) is easily found\footnote{
{For convenience,  $a_{0,1,2}$  can be found below,
for  $0<\xi\ll \eta\leq 1$ and $\eta^2\leq (d-2)(d-3)\alpha^2 q^2$:
\bea\label{solution}
&& \!\!\!\!  a_0^{d/2}=\frac{16 (d-3)}{\eta^2 (d-2)}\,\frac{a_0^2}{a_1^2},\qquad
\frac{a_0}{a_1}=\frac14 [-1\pm \sqrt{1+16\kappa}],\qquad
  \frac{a_2}{a_1}=\frac{1}{4}\alpha q (d-2) [-1\pm \sqrt{1-z}],\qquad\qquad
  \\
  \textrm{with}\!\!\!\!\!\!\!\!
  &&\! \!
  \kappa=\frac{d-2}{16\, (d-1)}\Big[\frac{d-1}{d-3} \frac{\eta^2}{4!\,\xi^2}-1\Big],\quad
  z= \frac{\eta^2}{(d-2)(d-3) \alpha^2 q^2}; \qquad\textrm{we have}\quad
  \vert \kappa\vert\gg 1, \, \vert z\vert<1\,\,
  \nonumber
  \eea
Solutions with both $\pm$ are valid;
  we also have $\vert a_{1,2}/a_0\vert\ll 1$, (convergent expansion).}}
 and then actions (\ref{S3}), (\ref{WAd}) also have 
  the same couplings, in $d$ dimensions. }

Action (\ref{WAd}) in $d$ dimensions was first introduced in \cite{DG1}
(for a review \cite{review}) as a natural regularisation 
of  action (\ref{WA})  (with $\hat R$ as regulator) in order to
respect Weyl gauge symmetry at quantum level, relevant for studying Weyl anomaly;
here this regularisation gains independent  mathematical support
from a more  general Weyl-DBI action.

Therefore, the Weyl-DBI action in $d$ dimensions (\ref{dbi}) generalises  the Weyl quadratic
gravity action  and  provides to it
an automatic analytical continuation to $d\!=\!4-2\epsilon$ while respecting
gauge symmetry (\ref{WGS}), as required for a gauge theory.
All fields are of geometric origin, with no added matter, Weyl scalar field
compensator or UV regulator, etc.

Action (\ref{dbi}) is very special among gauge theories: it is gauge invariant  with
dimensionless couplings in arbitrary $d$ dimensions,  hence it has no
need for a UV regulator, be it an extra scalar field \cite{Englert},
higher derivative operator or
a DR subtraction scale $\mu$ (scale demanded by  analytical continuation in usual
gauge theories). The analytical continuation of the Weyl-DBI action
is trivial, simply replace $d=4\rightarrow d=4-2\epsilon$ in  action (\ref{dbi}), with no
other change!
As a result, the gauge symmetry is  maintained at quantum level, also when coupled to matter
in a Weyl gauge invariant way, see  \cite{DG1} for an example.
For this reason one can argue that the
Weyl-DBI action is Weyl anomaly-free; this  is also supported by the fact that 
the leading order  of its expansion, eqs.(\ref{S3}) and (\ref{WAd}) i.e.
Weyl quadratic gravity   is itself  Weyl anomaly-free (when coupled to matter
in a Weyl gauge invariant way) \cite{DG1}.

\vspace{1.4cm}
\noindent {\bf $\bullet$ Weyl - DBI action in $d=4$ dimensions}

\bigskip\noindent
In the limit $d=4$ action (\ref{S3}) becomes
\medskip
\bea\label{S4}
S_4&=&\int d^4\sigma
\Big\{-
  \det\,[a_0 \,\hat R\, g_{\mu\nu}+a_1 \,\hat R_{\mu\nu}+ a_2 \,\hat F_{\mu\nu}]\,\Big\}^\frac12
\nonumber\\
&=&
\int d^4\sigma \sqrt{g}\, \Big[\,
  \frac{1}{4! \,\xi^2}  \,\hat R^2
  -\frac{1}{\zeta}
  \hat F_{\mu\nu} \, \hat F^{\mu\nu}
-\frac{1}{\eta^2}
  \hat C_{\mu\nu\rho\sigma}\hat C^{\mu\nu\rho\sigma}
  +\frac{1}{\eta^2}\, \hat G +\cO(X^3)\Big]
\eea

\medskip\noindent
where we denoted
\medskip
 \bea\label{couplings4}
 \frac{1}{4!\, \xi^2}=
 a_0^2+ \frac12 \,a_0\,a_1+ \,\frac{1}{24}\,a_1^2,\qquad\qquad
\frac{1}{\eta^2}=  \frac18\,a_1^2, \qquad\qquad
\frac{1}{\zeta}=-\frac14 \, a_2 (a_2+a_1 \alpha\,q).
\eea
One can express $a_i$, $i=0,1,2$ in terms of $\xi$, $\eta$, $\alpha$
as shown in (\ref{solution}) for  $d=4$, $\zeta=4$,
for perturbative couplings in the quadratic gravity action,
$\xi\ll \eta\leq \alpha\,\leq 1$, as
discussed earlier (after eq.(\ref{la})).
The constraint of a convergent expansion
$\vert a_{1,2}\vert\ll \vert a_0\vert$ is respected.

Action (\ref{S4}) is identical to (\ref{WA})  for this solution for
  $a_{0,1,2}$; the Euler term $\hat G$ is a total derivative if
  $d=4$ so it can be ignored in both (\ref{S4}) and (\ref{WA}).

Unlike the leading order i.e. Weyl quadratic gravity action, 
the exact DBI action can in principle have  more general values of
the couplings $a_i$, $i=0,1,2$, not restricted  by  convergence constraints of
its expansion, perturbativity, etc, in order to be physical.

As a result of (\ref{S4}), the Weyl-DBI action  inherits, in the leading order,
all the nice properties of Weyl quadratic gravity as a gauge theory,
mentioned in the introduction. 
Aside from the $C_{\mu\nu\rho\sigma}^2$ term\footnote{The effect of this term in the
action was extensively studied in \cite{Mannheim,Mannheim2}.} which is also
 present in Riemannian-based gravity theories\footnote{If not included classically,
 it is generated anyway at the quantum level. },
 Einstein-Hilbert gravity is recovered from action
 (\ref{S4}) in its broken phase shown in (\ref{EP}), after decoupling
 of massive $\w_\mu$; thus, Einstein-Hilbert gravity is also a broken phase of  the exact
 Weyl-DBI gauge theory, first line of (\ref{S4}).
This is an interesting result.

Let us  discuss the   terms $\cO(X^3)$ and higher in\,the\,action.
They bring in corrections like
\smallskip
\bea\label{qc}
\frac{\sqrt{g}}{\hat R} \hat R^{\mu\sigma} \hat R_{\sigma\rho} \hat R^{\rho}_{\,\,\,\,\mu},
\qquad
\frac{\sqrt{g}}{\hat R^2} (\hat C_{\mu\nu\rho\sigma}^2)^2,\qquad
{\rm etc,}
\eea

\medskip\noindent
The second term  is generated in order  $\cO(X^4)$.
Such terms are Weyl gauge
invariant and usually have  a non-perturbative interpretation; they
can be  important for a small Weyl scalar curvature  $\hat R$
or when the rapid convergence criterion  $\vert a_{1,2}/a_0\vert\ll 1$ is not respected.
The exact Weyl-DBI action sums up all  such terms.

These are apparently non-perturbative corrections to Weyl quadratic gravity; however they can
be generated   at quantum level by perturbative methods in a regularisation
and renormalization that respect the Weyl gauge symmetry 
(as they should, since this is a (quantum) gauge symmetry!).
For an example of such regularisation see eq.(\ref{S3})
provided by the Weyl-DBI action and also \cite{Englert,Misha1,DG1}.
Let us detail.
In order to preserve this symmetry at the quantum level, the 
usual subtraction scale $\mu$ of the dimensional regularisation (DR)
scheme is replaced by a field $\phi$ (``dilaton'' or would-be-Goldstone of Weyl gauge symmetry)
as in \cite{Englert,Misha1} or directly by the scalar curvature
$\hat R$ in our case  of eqs.(\ref{WAd}), (\ref{S3}) - notice that on the
ground state one actually has $\phi^2=-\hat R$, ($\hat R<0$), see text after
eq.(\ref{WA}). The scale $\mu$  is then generated  by the
vev of $\phi$  and then the (quantum) symmetry is broken only spontaneously!
The result  is that a series of (Weyl invariant) higher dimensional non-polynomial
operators is generated, suppressed by  powers of dilaton $\phi$
\cite{Misha1,Misha2,reg1,reg2,reg3}, which in our case correspond to terms
suppressed by powers of $\hat R$ which acts as a  regulator field here.

Such an approach  applied to action (\ref{WAd}) together with invariance under (\ref{WGS})
can then  generate, at perturbative quantum level, (non-perturbative)
non-polynomial terms like $\sqrt{g}\, (\hat C_{\mu\nu\rho\sigma}^2)^2/\hat R^2$ and similar.
Then the correction terms $\cO(X^3)$ and beyond, eq.(\ref{qc}), in the Weyl-DBI action, with
a  structure dictated only by symmetry (\ref{WGS}), are  similar 
to the (non-polynomial) quantum corrections to action (\ref{WA}) regularized as
in (\ref{WAd}). The classical Weyl-DBI action  thus captures non-perturbative
quantum corrections to Weyl quadratic gravity action, eq.(\ref{WA}).
This is an interesting result.

\vspace{0.8cm}
\noindent {\bf $\bullet$ Weyl - DBI action and conformal gravity}

\bigskip\noindent
Consider a special limit of the Weyl-DBI action in $d=4$ dimensions, eq.(\ref{S4}).
Assume that initially the Weyl gauge field is a ``pure gauge'' field.
In Weyl quadratic gravity, $\w_\mu$ is pure gauge when the Weyl gauge current
vanishes \cite{non-metric,GH}. Then its field strength
vanishes; formally this means $a_2=0$ in the exact Weyl-DBI action.
The action becomes
\bea
S_4&=&\int d^4\sigma
\Big\{-
  \det\,[a_0 \,\hat R\, g_{\mu\nu}+a_1 \,\hat R_{\mu\nu}]\,\Big\}^{\frac12}
\nonumber\\
&=&
\int d^4\sigma \sqrt{g}\, \Big[
  \frac{1}{24 \xi^2}  \,\hat R^2-\frac{1}{\eta^2}
  \hat C_{\mu\nu\rho\sigma}\hat C^{\mu\nu\rho\sigma}
+\frac{1}{\eta^2} \,\hat G  +\cO(X^3)\Big]
\eea

\medskip\noindent
This action simplifies when going to the Riemannian picture of the broken phase
in Einstein gauge/frame which is the physical one\footnote{In Riemannian picture the
action contains algebraic $\w_\mu$ dependence which is integrated out, see next.}.
One first linearises the term $\hat R^2$ in $S_{\bf w}$,
as explained earlier (text after eq.(\ref{WA}))
by introducing the scalar field $\phi$, then expresses
$\hat R$ in terms of its Riemannian notation, eq.(\ref{Rs}); the Weyl
tensor term does not change when going to the Riemannian picture, see (\ref{tc}),
so it does not affect the calculation when  $\w_\mu$ is integrated out
(similar for $\hat G$). After integrating $\w_\mu$  one finds 
in a Riemannian picture \cite{non-metric} (Section 3.1)
\bea\label{s33}
S_4=\int d^4\sigma \sqrt{g}\,
\Big\{ -\frac{1}{2\xi^2}\, \Big[\frac16 \phi^2 \,R+g^{\mu\nu} \partial_\mu\phi\partial_\nu\phi\Big]
- \frac{1}{4!\,\xi^2}\,\phi^4-\frac{1}{\eta^2}
   C_{\mu\nu\rho\sigma} C^{\mu\nu\rho\sigma}  +\cO(X^3)\Big\}.
  \eea

  \medskip
 We obtained a dilaton action  coupled to conformal gravity \cite{Mannheim}.
 The dilaton part of the action was  not added here by hand but has geometric origin in the
 $\hat R^2$ term, which is interesting and unlike in conformal gravity
 \cite{Kaku} where the dilaton part of the action is absent.
 Action (\ref{s33}) has local Weyl symmetry only  ($\w_\mu=0$ or pure gauge)
 and is  a particular limit of Weyl quadratic gravity eq.(\ref{S4}) and
 of its DBI version, which are  more general.
 This is understood from the fact that
 conformal gravity is not a {\it true} gauge theory of
 the full conformal group (with physical/dynamical gauge bosons).
 As a gauge theory of the conformal group, the conformal gravity action  cannot
 have kinetic terms for  special conformal or gauged dilatations \cite{Kaku}.
 Hence  it is  recovered from Weyl quadratic gravity which is a
 gauge theory of the smaller Weyl group (of dilatations
 times Poincar\'e symmetry), in the limit of vanishing gauge kinetic term
 for $\w_\mu$ ($\w_\mu$ pure gauge or zero) considered here.

 Finally, when $\phi$ acquires a vev in (\ref{s33}), one is in the Einstein frame/gauge and
  the Einstein-Hilbert term is generated and the dilaton decouples\footnote{
  unlike in (\ref{EP}) where it is eaten by $\w_\mu$.}. This can be seen
  by simply replacing $\phi\ra \langle\phi\rangle$ in (\ref{s33}); the first term in this
  equation recovers the Einstein-Hilbert term, the second decouples
  while the third term generates the cosmological
  constant. One then obtains an action similar to eqs.(\ref{EP}), (\ref{la}) but without
  the Proca action  of $\w_\mu$.

\vspace{0.5cm}
\noindent{\bf $\bullet$ Weyl - DBI action and U(1)}

\bigskip\noindent
So far we considered an analogue of the DBI action for the Weyl gauge dilatation
symmetry in conformal geometry. But one can also  consider  an additional
$U(1)$ gauge symmetry, then
\bea\label{Sdp}
S_4'=\int d^4\sigma\Big\{ -
  \det\,[a_0 \,\hat R\, g_{\mu\nu}+a_1 \,\hat R_{\mu\nu}+ a_2 \,\hat F_{\mu\nu}+a_3 \,\hat
    F_{\mu\nu}^y\,]\,\Big\}^\frac12
\eea
with  $\hat F^y_{\mu\nu}$ a $U(1)$ gauge field strength and $a_3$ a dimensionless constant.
An extension to $d$ dimensions is immediate.
Each term  under the determinant is Weyl gauge invariant. Then
\be
S_4'=
S_4+
\int d^4\sigma \sqrt{g}\, 
\Big[- \zeta_1  \hat F^y_{\mu\nu} \,\hat F^{y\,\mu\nu} 
  -\zeta_2\,\hat F_{\mu\nu}\,\hat F^{y \mu\nu}+\cO(X^3)\Big]
\ee
%
with $S_4$ as in eq.(\ref{S3}) for $d=4$ and
\bea
\zeta_1&=& -\frac14\, a_3^2\,
\qquad
\zeta_2=-\frac12\, a_3\,
\Big(a_2+a_1\, \alpha\,q/2\,\Big).
\eea
%
With $\zeta_1>0$ for a well defined gauge kinetic term
of $U(1)$,  then $a_3$ must be  imaginary, then $\zeta_2$ is also imaginary,
for real $a_{1,2}$. This situation does not change if we set $a_2=0$ or in the 
limit of integrable Weyl geometry (when $\w_\mu$ is a pure gauge field, $\hat F_{\mu\nu}=0$).
$S_4^\prime$  must thus be amended by a hermitian conjugate in the
rhs of eq.(\ref{Sdp})\footnote{A related  DBI-like action could be 
 $S=\int d^4\sigma\,\{
   -\det [
    a_0 \, \delta^a_b \,\hat R \,g_{\mu\nu}+
     a_1 \hat R^a_{\,\,\,b\,\mu\nu}
     +a_2 \delta^a_{b}\, \hat F_{\mu\nu}^y ]\,\}^{1/2}$;
   with a  trace understood over the tangent-space indices $a, b$; this is Weyl gauge invariant;
 the calculation is  similar.}.

Under suitable assumptions a DBI action can be seen as a low energy effective description
of a D-brane action in string theory, so one could ask,
somewhat naively, how close a Weyl - DBI action
  like (\ref{Sdp}) or (\ref{S4}) 
 is   to  a $D_3$-brane action \cite{Tong} in the background of
 closed string modes $G_{\mu\nu}$, two-form $B_{\mu\nu}$ and dilaton $\Phi$.
 Weyl gauge invariance is not a symmetry in strings, the brane tension/$\alpha'$ break it.
 But not all hope is lost,  some similarities still exist:
  consider the $D_3$ brane action   in the background mentioned.
  The brane tension/$\alpha'$ is ultimately generated  by the dilaton;
  similarly, in Weyl geometry the dilaton propagated by $\hat R^2$ (in the
 expanded action) generates the  Planck scale instead, eq.(\ref{la}). Factorising
 $\hat R^2$ in front of  $\det$  in (\ref{Sdp}) and using an equation of motion
to replace it by $\langle\phi\rangle^4$  (recall $\phi^2=-\hat R$ in leading order  $\cO(X^3)$),
would seem to bring this action closer to a $D_3$-brane action with the brane tension replaced
by the dilaton $\cT_3\sim\langle\Phi\rangle^4$. Further,
in the $D_3$ brane action  the anti-symmetric $B_{\mu\nu}$ `combines'
with the field strength of $U(1)$, while its field strength $H=dB$  plays the
role  of an anti-symmetric  torsion  {\it tensor};
in  conformal geometry a  counterpart to $B_{\mu\nu}$ could be
$\hat F_{\mu\nu}$ (of $\w_\mu$) which can also  mix with  the field strength
of $U(1)$ while respecting Weyl gauge invariance,  but one can show that
torsion is here {\it vectorial} only \cite{CDA}; one must thus go
beyond such assumption and consider in Weyl conformal geometry a
totally anti-symmetric torsion tensor. It is worth exploring 
this  relation in some detail.

\section{Conclusions}

We constructed  the analogue of a DBI action in  conformal geometry in
$d$ dimensions. For this we used a  {\it Weyl gauge covariant} formulation
of conformal geometry in $d$ dimensions, suitable for a gauge theory,
in which this geometry is {\it metric}. We found a  general Weyl-DBI
theory of gravity with Weyl gauge symmetry in arbitrary $d$ dimensions.
This  theory is very special among gauge theories in that its action is Weyl gauge invariant
with {\it dimensionless} couplings for any  dimension $d$; hence the action
has {\it no need for a UV regulator} (be it an extra scalar field, higher
derivative operator or DR subtraction scale) necessary in common gauge theories
when the theory is analytically continued from $d=4$  to $d=4-2\epsilon$.
Here the analytical continuation is trivial, just replace $d=4$ by
$d=4-2\epsilon$, or by any $d$\, with no other change in the Weyl-DBI action!
Its series expansion shows that the Weyl scalar curvature $\hat R$ plays 
the role of the UV regulator,  while preserving the Weyl gauge symmetry
of the action. For this reason, with gauge symmetry manifest in $d$ dimensions
one can say that, when coupled to matter in a Weyl gauge invariant way,
the Weyl-DBI gauge theory is Weyl anomaly-free. This  is also supported
by the fact that  the leading order of its expansion i.e. Weyl quadratic gravity
is Weyl-anomaly free.

The exact Weyl-DBI action  naturally extends  the general Weyl quadratic gravity
action which is itself a gauge theory of the Weyl group.
Indeed, for $d=4$ the leading order of an expansion of the Weyl-DBI action
becomes the  Weyl quadratic gravity action which is physically relevant:
the Weyl gauge  symmetry is broken by Stueckelberg mechanism and one
recovers the  Einstein-Hilbert gravity in the broken phase, with
cosmological constant $\Lambda>0$.
However, the exact  Weyl-DBI action is more general - it is also valid 
for e.g.  small Weyl scalar curvature which affects the convergence of
the expansion. The symmetry  breaking is in a sense geometric, since
there are no matter fields or Weyl scalar compensators added ``by hand''
to this purpose, in these actions.

For $d$ dimensions, the Weyl-DBI action  in the leading order of its series expansion
gives an analytical continuation to $d\!=\! 4-2\epsilon$ of Weyl quadratic gravity
that remains Weyl gauge invariant at quantum level. This gives mathematical support
to using such Weyl gauge invariant regularisation in $d=4$ Weyl quadratic gravity,
as already done in \cite{DG1,review}.
All the remaining,  apparently non-perturbative higher order corrections in
the expansion, given by $\cO(X^3)$ and beyond,
have a non-polynomial form and are  similar to
those generated perturbatively by   quantum corrections in $d=4$ Weyl
quadratic gravity with such Weyl gauge invariant regularisation,
(e.g. $(C_{\mu\nu\rho\sigma}^2)^2/\hat R^2$). The classical  Weyl - DBI action thus
captures (gauge invariant) non-perturbative quantum corrections. This is an important
result.

In so-called ``integrable geometry'' limit i.e. when the Weyl gauge field  is
``pure gauge'' (i.e. non-dynamical)
the Weyl-DBI action becomes in the leading order, for $d=4$,
the usual conformal gravity action  plus  a dilaton action
with local Weyl symmetry. Hence conformal gravity (usually regarded as
a gauge theory of the full conformal group\footnote{but with no
associated physical/dynamical gauge bosons of
special conformal and dilatation symmetries!})
is actually just
a particular limit of  Weyl quadratic gravity (of smaller
Weyl gauge group!) and of its generalisation into the
Weyl-DBI action. These interesting results deserve further study.

\vspace{1.5cm}
\noindent
{\bf Acknowledgements:\,\,\,}
The author thanks C. Condeescu and A. Micu  for many interesting
discussions on Weyl conformal geometry.

\end{document}